# Phase transitions in voting simulated by an intelligent Ising model

*Guanyu Xu[1], Jiahang Chen[1], Xin Zhou[1,2], Yanting Wang[1,2,3,]\**

[1]School of Physical Sciences, University of Chinese Academy of Sciences, Beijing 100049, China

[2]Wenzhou Institute, University of Chinese Academy of Sciences, Wenzhou 325001, China

[3]Institute of Theoretical Physics, Chinese Academy of Sciences, Beijing 100190, China

**Email:** wangyt@itp.ac.cn

**Abstract**

Voting is an important social activity for expressing public opinions. By conceptually considering a group of voting agents to be intelligent matter, the impact of real-time information on voting results is quantitatively studied by an intelligent Ising model, which is formed by adding nonlinear instantaneous feedback of the overall magnetization to the conventional Ising model. In the new model, the interaction strength becomes a variable depending on the total magnetization rather than a constant, which mimics the scenario that the decision of an individual during vote influenced by the dynamically changing polling result during the election process. Our analytical derivations along with Mote Carlo simulations reveal that, with a positive feedback, the intelligent Ising model exhibits phase transitions at any finite temperatures, a feature lacked in the conventional one-dimensional Ising model. In all dimensions, by varying the feedback strength, the system changes from going through a second-order phase transition to going through a first-order phase transition with increasing temperature, and the two types of phase transitions are connected by a tricritical point. This study on the one hand demonstrates that the intelligent matter with a nonlinear adaptive interaction can exhibit qualitatively different phase behaviors from conventional matter, and on the other hand shows that, during voting, even unbiased feedback may possibly induce spontaneous symmetry breaking, leading to a biased outcome where one side of the vote becomes favored.

**Introduction**

Voting is an essential mechanism of modern democratic life for expressing public opinion (1). The intent of voting is for the outcome to reflect the will of the majority of participants, but in the information era, media and social platforms may potentially influence or even manipulate voting results through opinion guidance (2, 3). It is therefore necessary to quantitatively study the impact of real-time information on voting outcomes with the aid of statistical physics (4–6). In this study, we consider a form of "fair" real-time information feedback, where the media provides unbiased information: voters are allowed to know the current overall voting trend in real time and make their next voting decisions based on this information.

To derive general conclusions, we have to introduce a highly simplified statistical model. We consider only binary voting decisions, where each agent can vote either "yes" or "no". The Ising model is well-suited for simulating such scenarios. In the Ising model, each site (representing a voter) has a spin variable that can be +1 (spin up, representing a "yes" vote) or –1 (spin down, representing a "no" vote). Each agent is influenced by its nearest neighbors ($z = 2d$ neighbors in a $d$-dimensional model), tending to align its vote with those neighbors. The Hamiltonian (energy function) of the conventional Ising model without external fields (7) is

$$H = -J \sum_{\langle i,j \rangle} s_i s_j, \qquad (1)$$

where $s_i = \pm 1$ is the spin (vote) of agent $i$, the sum runs over all pairs of neighboring sites $\langle i, j \rangle$, and the constant $J > 0$ is the interaction strength favoring the alignment of neighboring spins. The periodic boundary condition is applied, so the last and first spins in one dimension are a pair of neighbors. The system temperature $T$ is a relative concept compared to the interaction strength: a higher temperature corresponds to weaker effective influence of neighbors, meaning that voters make more independent decisions (higher "autonomy"), whereas a lower temperature means neighbors have a stronger influence (voters are more conformist). The total magnetization $M = \sum_{i=1}^{N} s_i$ represents the net vote count (difference between "yes" and "no" votes) and $N$ is the number of sites (voters).

The conventional Ising model is a well-studied paradigm of phase transitions in statistical physics. In one dimension (a linear chain of spins), the Ising model does not exhibit a phase transition at finite temperatures – thermal fluctuations are strong enough to prevent any long-range order (8). In other words, for the 1D Ising model, the ensemble-averaged magnetization $\langle M \rangle = 0$ remains zero for any $T > 0$, indicating that without external influence, the numbers of "yes" and "no" votes tend to balance out. By contrast, in two and higher dimensions, the Ising model undergoes a second-order phase transition at a critical temperature $T_c$. Above $T_c$, the system is disordered: thermal fluctuations dominate, and the average magnetization is zero (no overall majority). Below $T_c$, the system becomes ordered: the interactions dominate over thermal noise, leading to a spontaneous magnetization (a majority of "yes" or "no" votes). The critical temperature marks the onset of collective behavior where the population can polarize into a majority opinion when the dimension $d > 1$, but not when $d = 1$.

We are now interested in the question: How does real-time knowledge of the overall voting trend affect the outcome? To investigate this problem, we modify the conventional Ising model to include positive feedback in the interaction between neighbors. In our *intelligent Ising model*, the interaction strength between neighbors is not a constant anymore; instead, it increases with the instantaneous total magnetization, creating a positive feedback loop. This mimics a scenario where voters, seeing a growing strength of majority, feel increased pressure to conform with their neighbors. The Hamiltonian for the intelligent Ising model can be written as

$$H = -J(m) \sum_{\langle i,j \rangle} s_i s_j , \tag{2}$$

where the interaction becomes a function of the specific magnetization $m \equiv M/N$:

$$J(m) = 1 + km^2 , \tag{3}$$

where $k$ is the coupling strength between $J$ and $m$.

The concept of the intelligent Ising model draws inspiration from recent advances in soft matter physics (9). Soft materials (such as polymers, liquid crystals, colloids, and biological matter) owe their "softness" to a delicate balance to the free energy landscape between enthalpic and entropic contributions (10). Building on this, the field of active matter has emerged to describe systems of many self-driven units (11, 12) (e.g. bacterial swarms, bird flocks, fish schools) that consume energy to move or make decisions. Active matter systems can exhibit collective behaviors (like flocking or swarming) (13, 14) that are not seen in equilibrium systems, due to the continuous input of energy at the individual level. However, traditional active matter models do not explicitly include adaptive feedback in interactions (15). We propose that an essential feature of *intelligent agents* (such as human voters (16, 17) or animals with learning) is the ability to adaptively adjust interactions in response to dynamic conditions via feedback mechanisms (18–20). In other words, unlike passive particles or simple self-propelled agents, intelligent agents can change how strongly they influence or are influenced by others based on input information. From the viewpoint of statistical mechanics, a group of intelligent agents can be named as *intelligent matter*, which can be used to model and study a collective of behaving individuals with intelligence. This is precisely what our intelligent Ising model implements: the interaction strength adapts in response to the global magnetization (the current vote tally).

In this work, we study the phase behaviors of the intelligent Ising model using both analytical derivations and computer simulations. Our findings reveal striking differences from the conventional model due to the feedback mechanism: A phase transition occurs even in one dimension as the temperature is varied, and a small imbalance can grow and lock in due to the increasing coupling, leading to a spontaneous magnetization at low temperatures; A second-order phase transition exhibits at a weaker coupling strength $k$ and a first-order phase transition at a stronger $k$; A special point called the tricritical point (8) marking the boundary where the transition changes its order from second to first. This rich phase behavior arises purely from the feedback-induced adaptation of interactions. Translating these results back to the voting scenario, we demonstrate that real-time information about collective opinion can significantly alter voting dynamics. Even unbiased feedback may possibly induce spontaneous symmetry-breaking, leading to a biased outcome where one side of the vote becomes favored.

**Results**

*Analytic Solution of the 1D Intelligent Ising Model*

To analyze the phase behavior of the model, we utilize the free energy landscape

$$F(m) = -\frac{1}{\beta} \ln Z_N(m), \tag{4}$$

where $\beta \equiv \frac{1}{k_B T}$ with $k_B$ the Boltzmann constant, and the partition function corresponding to Eq. (2) is

$$Z_N(m) = \sum_{\{s_i\}} \delta\left(\frac{1}{N}\sum_{i=1}^{N} s_i - m\right) \exp\left(\beta J(m) \sum_{i=1}^{N} s_i s_{i+1}\right). \tag{5}$$

In one dimension, similar to the conventional Ising model (7), this model has an exact analytical solution. As detailed in the Methods section, we obtain the exact analytical solution of the free energy landscape

$$\begin{aligned} f(m) = & -T \ln\left[\sqrt{e^{2J(m)/T} + e^{-2J(m)/T} \frac{m^2}{1-m^2}} + \frac{e^{-J(m)/T}}{\sqrt{1-m^2}}\right] \\ & + mT \ln\left[e^{-2J(m)/T} \frac{m}{\sqrt{1-m^2}} + \sqrt{1 + e^{-4J(m)/T} \frac{m^2}{1-m^2}}\right] \end{aligned}. \tag{6}$$

This expression characterizes the thermodynamic behavior of the system by the fact that its minima corresponding to stable or metastable phases, and its phase behavior is determined by the feedback strength $k$ and the temperature $T$.

The exact analytical solution of the 1D intelligent Ising model reveals rich phase behavior modulated by the coupling strength $k$. The phase diagram, as shown in Figure 1(a), is plotted in the $(k, T^*)$ plane, where the reduced temperature $T^* = \frac{zJ}{k_B}$. The white region corresponds to the paramagnetic phase characterized by $\langle m \rangle = 0$, while the blue region represents the ferromagnetic phase with spontaneous magnetization where $\langle m \rangle \neq 0$. The dashed curve marks the spinodal line, beyond which the metastable states vanish.

From the phase diagram, we observe that at $k = 0$, the critical temperature $T_c$ approaches zero, consistent with the well-known result of the conventional 1D Ising model. As $k$ increases from zero, $T_c$ rises sharply, indicating that even a small nonlinear feedback strength $k$ can induce a finite-temperature phase transition in the 1D Ising model. For larger $k$, the growth of $T_c$ continues but becomes more gradual.

Interestingly, the nature of the phase transition also changes with $k$. For $k < k_c$, the phase transition from low to high temperature is continuous (second-order). By contrast, for $k > k_c$, the phase transition becomes discontinuous (21) (first-order) with hysteresis. This change in transition order is clearly reflected in the free energy landscapes: as shown in Figure 1(b), for $k = 0.1$, the minimum of the free energy landscape evolves continuously with temperature, indicating a continuous phase transition; while for $k = 0.4$, Figure 1(c) reveals the emergence of competing minima separated by a barrier, signaling metastability and phase coexistence. These features suggest the presence of a tricritical point on the phase boundary where the transition shifts from second to first order.

To accurately locate the tricritical point, we perform a Landau expansion of the free energy up to the sixth-order term in the order parameter (8) as

$$f(m;k,T) = f_0 + a(k,T)m^2 + b(k,T)m^4 + c(k,T)m^6 + O(m^7). \qquad (7)$$

When $a(k,T) = b(k,T) = 0$ and $c(k,T) > 0$, the system is at the tricritical point with $k_c = 1.115$ and $T_c = 1.077$. This point marks the boundary between the continuous (second-order) and discontinuous (first-order) phase transitions, indicating a fundamental change in the nature of the phase transition.

### *Simulation Results for the 1D Intelligent Ising Mode*

To validate the analytical predictions, we employ the Metropolis Monte Carlo algorithm (22) to simulate the evolution of the spin configurations in the 1D intelligent Ising model at various values of $k$. The results are presented in Figure 2. For $k = 0.1$, both the energy per spin (Figure 2(a)) and the magnetization per spin (Figure 2(b)) change continuously as the system is heated up or cooled down across the transition region. The lack of hysteresis confirms the second-order nature of the phase transition at a low $k$. By contrast, abrupt jumps in both energy and magnetization are observed near the critical temperature at $k = 0.2$, indicating phase coexistence during the transition characterizing a first-order phase transition. These simulation results are in excellent agreement with the theoretical predictions derived from the exact solution.

### *Simulation Results for the 2D and 3D Intelligent Ising Models*

Building on the one-dimensional analysis, we extended our study to higher-dimensional systems by Monte Carlo (MC) simulation. Figure 3 shows the simulated phase diagram of the 2D intelligent Ising model. In the limit $k \to 0$, the critical temperature $T_c$ approaches that of the conventional 2D Ising model (23), and $T_c$ increases when $k$ becomes larger.

Figure 4 displays the temperature dependence of the energy and magnetization per spin for various values of $k$. At a small $k$, both quantities vary smoothly with temperature, consistent with a continuous phase transition. As $k$ increases, the transition becomes sharper and discontinuity emerges, signaling the crossover to a first-order-like behavior. Therefore, analogous to the 1D case, a tricritical point connecting the two kinds of phase transitions exists.

We performed the same simulations and analysis for the 3D intelligent Ising model. The resulting phase diagram, shown in Figure 3(b), exhibits similar behavior: The critical temperature increases with $k$, and in the limit $k \to 0$, the transition temperature approaches that of the conventional 3D Ising model (23). Figure 4 presents the temperature dependence of the energy and magnetization per spin for different $k$ values. As in the 2D case, smooth variations are observed for small $k$ values, and a sharp transition and discontinuity appearing at large $k$ values, again supporting the existence of a tricritical point in the 3D system.

*Mean-Field Solution*

The mean-field theory provides further insight into the phase behavior of the intelligent Ising model across dimensions. Under the mean-field approximation, we substitute the mean-field decomposition $s_i = \phi + \delta s_i$ into the system's Hamiltonian, Eq. (2), where $\phi = \langle s_i \rangle$ is the average magnetization and $\delta s_i$ denotes the fluctuations. By assuming that the fluctuations are small, we can neglect the second-order and higher-order terms of $\delta s_i$. As detailed in the Methods section, the resulting free energy landscape can be written as

$$f(\phi) = \frac{z^2}{2}\phi^2\left(1+3k\phi^2\right) - k_B T \ln\left[2\cosh\left(\beta z\left(\phi + 2k\phi^3\right)\right)\right]. \tag{8}$$

Figure 7(a) presents the phase diagram in the $(k, T^*/z)$ plane derived from the mean-field theory. The white region indicates the paramagnetic phase ($\langle m \rangle = 0$), the blue region represents the ferromagnetic phase ($\langle m \rangle \neq 0$), and the dashed line is the spinodal line. Expanding the free energy landscape near $\phi = 0$, we obtain

$$f(\phi) = f_0 + a(k,T)\phi^2 + b(k,T)\phi^4 + c(k,T)\phi^6 + O(\phi^7). \tag{9}$$

The tricritical point can thus be determined by letting $a(k,T) = b(k,T) = 0$, yielding $k_c = \frac{1}{6}, T_c = z$.

Figures 7(b) and 7(c) illustrate the free energy landscapes at different values of the coupling parameter $k$. At $k = 0.1$, the system exhibits a continuous phase transition; while at $k = 0.4$, the phase transition becomes discontinuous, clearly demonstrating the presence of a tricritical point.

Although the phase diagram shows similar tricritical behavior for the mean-field model with the finite-dimensional models, there is a key difference that the continuous phase transition temperature $T_c$ of the mean-field model does not depend on $k$, which might be attributed to the fact that the mean-field approximation greatly suppresses thermal fluctuations, especially around the critical points.

## Conclusions and Discussion

*Phase Transitions in the Intelligent Ising Models*

The study of the intelligent Ising model introduced in this work demonstrates how feedback mechanisms, realized through a magnetization-dependent coupling coefficient $J(m)$, can qualitatively change the phase behavior of the conventional Ising model. Both analytical derivation and MC simulation for the 1D system reveal the emergence of a tricritical point, where the nature of the phase transition changes from continuous (second-order) to discontinuous (first-order) as the nonlinear feedback parameter $k$ increases, in contrast to the phase behavior of the conventional 1D Ising model that no phase transitions appear at finite temperatures. In two and three dimensions, by MC simulation, we observe a similar phase transition behavior from continuous to discontinuous as $k$ increases. The mean-field theory confirms that the above phenomenon is dimension-independent, providing a generalized framework for understanding the tricriticality induced by the feedback mechanism in various dimensions.

In the conventional Ising model, the critical temperature increases with spatial dimension (24–26), and the critical behavior approaches the predictions of the mean-field theory as the dimension grows. We anticipate that the phase diagram in the $(k,T)$ plane will similarly converge toward the mean-field result in higher dimensions. In our current simulations, important effects such as finite-size corrections, which are known to play a significant role in the Ising model, have not been systematically addressed. Future studies could employ algorithms more suited to exploring phase transitions, such as the Wang–Landau method (27), in combination with finite-size scaling analysis to determine the phase boundaries more precisely (28).

*Sociophysical Interpretation: Opinion Polarization Induced by Poll Feedback*

In our model, the nonlinear coupling $J=1+km^2$ can be interpreted as a feedback mechanism induced by opinion polls during an election. As polling results become increasingly biased toward one side, the absolute magnetization $|m|$ increases, leading to a larger $J$, which enhances the tendency of individuals to align their opinions with their immediate social environment. Importantly, this does not imply a direct shift toward the poll-leading side, but rather a reinforcement of local consensus—a phenomenon similar to political homophily (29) or the echo chamber effect (30), where individuals become more influenced by surrounding opinions under perceived social pressure.

We consider a dynamic scenario where polling continues throughout the election period and individuals grow more responsive to these polls as the election day becomes closer (31, 32). This temporal increase in poll sensitivity is modeled in the intelligent Ising model by a gradual increase in the feedback strength $k$. Initially, assuming a balanced race with $m=0$, the system remains disordered. As $k$ increases, the system exhibits two qualitatively different behaviors depending on the temperature. At low temperatures, where fluctuations are relatively weak and individuals are more likely to retain being aligned with their local environment, the system undergoes a continuous phase transition: the order parameter $m$ grows smoothly from zero, signaling a gradual polarization of public opinion. By contrast, at high temperatures, where individual opinions are more independent and fluctuations are strong, the system remains in a disordered state as $k$ increases, until a critical threshold is reached. At this point, the system undergoes a discontinuous, first-order phase transition, and the magnetization $m$ jumps abruptly away from zero, indicating a rapid, avalanche-like shift in collective opinion.

This behavior suggests a sociopolitical analogy: in more collectivistic societies (corresponding to low temperatures), collective opinion emerges progressively as individuals respond to local consensus under polling pressure. In more individualistic societies (high temperatures), opinions remain dispersed until polling influence becomes strong enough to trigger an opinion cascades (33). Hence, our model captures two fundamentally different routes to consensus formation: smooth evolution and abrupt polarization, depending on the social temperature and feedback strength.

It is worth noting that the above interpretation is based on the analytical results of the 1D intelligent Ising model. In higher-dimensional systems, such as the 2D and 3D cases studied in our simulations, spontaneous symmetry breaking (34) occurs at low temperatures even without the feedback mechanism, leading to the emergence of collective opinion polarization. In these models, the system exhibits three qualitatively different regimes depending on temperature. At low temperatures, spontaneous ordering dominates; at intermediate and high temperatures, the behavior is similar to that observed in the 1D case, with the feedback strength $k$ controlling the onset and nature of polarization. This highlights the role of dimensionality in shaping the consensus formation process. In the Ising model, dimensionality reflects the extent of connectivity among individuals: Higher dimensions correspond to individuals being influenced by a larger number of neighbors. This increased local interaction plays a significant role in facilitating the emergence of collective opinion, making dimensionality an important factor in the dynamics of consensus formation.

In summary, we incorporated feedback interactions into the conventional Ising model to build up an intelligent Ising model, leading to the emergence of complex phase behaviors. This model shows how adaptive interactions (a hallmark of intelligent matter) can qualitatively change phase transition behavior – enabling order to emerge where it otherwise would not, and tuning the nature of the transition between continuous and discontinuous. Our findings underscore that, in modern democratic processes, the way that information is disseminated can have subtle but profound effects on the aggregation of preferences. Understanding these effects is crucial for designing fair voting systems and for interpreting election results in an age of instant communication and social media influence.

**Materials and Methods**

*Exact Solution of the 1D Intelligent Ising Model*

To analyze the phase behavior of our model, we first introduce the free energy landscape as a function of magnetization $m$:

$$f(m) = -k_B T \ln Z(m), \qquad (10)$$

where the partition function $Z(m)$ is defined as

$$Z(m) = \sum_{\{s_i\}} \delta\left(\frac{1}{N}\sum_{i=1}^{N} s_i - m\right) \exp\left[\beta J(m) \sum_{i=1}^{N} s_i s_{i+1}\right]. \qquad (11)$$

To evaluate $Z(m)$, we make use of the known solution for the conventional Ising model with an external magnetic field $h$ applied, whose Hamiltonian is

$$H = -J(m)\sum_{\langle i,j \rangle} s_i s_j + h\sum_{i} s_i. \qquad (12)$$

Its partition function is

$$Z_{\text{conv}} = \exp(N\beta J)\left(\cosh(\beta h) + \sqrt{\sinh^2(\beta h) + \exp(-4\beta J)}\right)^N. \tag{13}$$

To relate it to $Z(m)$, we employ a mathematical transformation as described below. Eq. (13) can be alternatively written as

$$Z_{\text{conv}}(\beta, h) = \int_{-1}^{1} dm \sum_{\{s_i\}} \delta\left(\frac{1}{N}\sum_{i=1}^{N} s_i - m\right) \exp\left(\beta J \sum_{i=1}^{N} s_i s_{i+1} + \beta h \sum_{i=1}^{N} s_i\right),$$
$$= \int_{-1}^{1} dm \exp\left[-\beta N f_{\text{conv}}(\beta, h, m)\right] \tag{14}$$

which can be applied with the saddle point approximation in the thermodynamic limit to replace the integral by its value at the most probable magnetization $m^*$ to obtain

$$Z_{\text{conv}}(\beta, h) \approx \exp\left[-\beta N f_{\text{conv}}(\beta, h, m^*)\right] \approx \sum_{\{s_i\}} \delta\left(\frac{1}{N}\sum_i s_i - m^*\right) \exp\left(\beta J \sum_{i=1}^{N} s_i s_{i+1}\right) \exp(\beta h N m^*). \tag{15}$$

Here, $m^*$ satisfies the equation of state for the conventional 1D Ising model under $h$:

$$\beta h(\beta, m^*) = \ln\left(\sqrt{\frac{m^*}{1-m^*}} e^{-2\beta J} + \sqrt{\frac{m^{*2}}{1-m^{*2}} e^{-4\beta J} + 1}\right). \tag{16}$$

Substituting the above expression for $\beta h$ into Eq. (15), we eliminate the dependence on the external field $h$ and obtain

$$\sum_{\{s_i\}} \delta\left(\frac{1}{N}\sum_{i=1}^{N} s_i - m\right) \exp\left(\beta J \sum_{i=1}^{N} s_i s_{i+1}\right)$$
$$= \left(\sqrt{e^{2\beta J} + e^{-2\beta J} \frac{m^2}{1-m^2}} + \frac{e^{-\beta J}}{\sqrt{1-m^2}}\right)^N \left(e^{-2\beta J} \frac{m}{\sqrt{1-m^2}} + \sqrt{1 + e^{-4\beta J} \frac{m^2}{1-m^2}}\right)^N. \tag{17}$$

Here we rewrite $m^*$ as $m$. This equation is purely mathematical, and remains valid when we generalize the coupling constant to be a function of magnetization, i.e. $J \to J(m)$. Therefore, the partition function of our feedback-regulated model becomes

$$e^{-\beta N f(m)} = Z(m)$$
$$= \left(\sqrt{e^{2\beta J(m)} + e^{-2\beta J(m)} \frac{m^2}{1-m^2}} + \frac{e^{-\beta J(m)}}{\sqrt{1-m^2}}\right)^N \left(e^{-2\beta J(m)} \frac{m}{\sqrt{1-m^2}} + \sqrt{1 + e^{-4\beta J(m)} \frac{m^2}{1-m^2}}\right)^N. \tag{18}$$

Taking the logarithm and being divided by $-\beta N$, we obtain the expression for the free energy landscape

$$f(m) = -T \ln\left(\sqrt{e^{2J(m)/T} + e^{-2J(m)/T}\frac{m^2}{1-m^2}} + \frac{e^{-J(m)/T}}{\sqrt{1-m^2}}\right)$$
$$+ mT \ln\left(e^{-2J(m)/T}\frac{m}{\sqrt{1-m^2}} + \sqrt{1 + e^{-4J(m)/T}\frac{m^2}{1-m^2}}\right) \quad . \tag{19}$$

This function forms the theoretical basis not only for this study, but also for future investigations of various intelligent matter models, including the identification of metastable states and the nature of phase transitions under different feedback mechanisms.

*Monte Carlo simulation*

To evolve the system toward thermal equilibrium, we adopted the conventional Metropolis algorithm with a key modification incorporating the feedback mechanism inherent to our Intelligent Ising model in our MC simulations. In this model, the coupling coefficient is not a constant, but instead depends dynamically on the instantaneous magnetization $m$ by introducing a global feedback loop. Specifically, we define the coupling as $J(m) = 1 + km^2$, where $k$ is the feedback strength.

At each MC step, a spin is randomly selected and proposed for flipping. To determine whether the flip should be accepted, we compute the corresponding energy change $\Delta E$ including both local and global contributions due to the feedback term. Let $m$ and $m'$ be the system magnetization before and after the flip, respectively, and $\Delta J = k(m'^2 - m^2)$ the resulting change in the coupling constant. The system energy before and after the flip can be respectively expressed as

$$E = -J \sum_{\langle i,j \rangle} s_i s_j \tag{20}$$

and

$$E' = -(J + \Delta J) \sum_{\langle i,j \rangle} s_i s_j + 2(J + \Delta J) s_k \sum_{j \in \text{nn}(k)} s_j, \tag{21}$$

where $\langle i,j \rangle$ denotes the summation over all nearest-neighbor spin pairs, and $\text{nn}(k)$ represents the set of nearest neighbors of the flipping spin $s_k$. The second term in $E'$ accounts for the change in interaction due to the flip of $s_k$, which only affects its local neighbors.

The energy difference before and after the flip $\Delta E = E' - E$ is then

$$\Delta E = \frac{\Delta J}{J} E + 2(J + \Delta J) s_k \sum_{j \in \text{nn}(k)} s_j. \tag{22}$$

This formulation avoids recomputing the full energy at every step and significantly improves computational efficiency, despite the involvement of a global coupling term through the feedback mechanism. The proposed flip is accepted with the probability $P = \min(1, e^{-\Delta E/T})$, which ensures that the Markov chain satisfies the detailed balance and converges to the correct Boltzmann distribution.

*Numerical Protocol*

MC simulations were performed in one, two, and three spatial dimensions. All systems were started from random configurations corresponding to a near-zero initial magnetization.

For the 1D case, we used a system size of $L=10^4$ and performed the simulations for $10^8$ MC steps, focusing on representative values of $k$, including 0.1 and 0.4. In two dimensions, we used square lattices of size $100 \times 100$, also with $10^8$ steps. The coupling parameter $k$ was varied from 0 to 0.5 with an increment of 0.05, and the temperature $T$ was swept from 2.0 to 4.5 with a finer interval when being closer to the transition region. The 3D simulations were conducted on a cubic lattice of size $20 \times 20 \times 20$, using the same $k$ values and a temperature range from 4.0 to 6.5. In all cases, the first 30% of the simulation steps were discarded for equilibration, and physical observables, including the average energy, magnetization, and their conventional deviations, were computed from the equilibrated data sampled after the equilibration. These quantities were used to characterize the phase behavior and identify the transition types under various parameter regimes.

*Mean-Field Solution*

In our model, the coupling coefficient $J$ depends on the average magnetization of the system, and the Hamiltonian is written as

$$H(\{s_i\}) = -(1+km^2)\sum_{\langle i,j \rangle} s_i s_j = -\left[1 + \frac{k}{N^2}\left(\sum_{i=1}^{N} s_i\right)^2\right]\sum_{\langle i,j \rangle} s_i s_j, \tag{23}$$

where $m = \frac{1}{N}\sum_{i=1}^{N} s_i$ is the average spin of the system.

In the mean-field approximation, each spin $s_i$ is decomposed into its average value $\phi = \langle s_i \rangle$ and a fluctuation term $\delta s_i$:

$$s_i = \phi + \delta s_i. \tag{24}$$

Assuming the fluctuation is small, we retain only the first-order terms in $\delta s_i$ and neglect higher-order contributions.

Expanding the term $\left(\sum_{i=1}^{N} s_i\right)^2$, we obtain

$$\left(\sum_{i=1}^{N} s_i\right)^2 = -N(N-1)\phi^2 + 2(N-1)\phi \sum_{i=1}^{N} s_i. \tag{25}$$

Substituting this into the expression for $J$, we get

$$J \approx 1 - k\phi^2 + 2k\phi m. \tag{26}$$

Considering that the total number of nearest-neighbor spin pairs is $\frac{zN}{2}$, the two-spin interaction term under the mean-field approximation becomes

$$\sum_{\langle i,j \rangle} s_i s_j \approx \frac{zN}{2}\phi^2 + z\phi \sum_{i=1}^{N} \delta s_i. \tag{27}$$

By combining the above expansions together, the Hamiltonian is reduced to a form containing only single-spin terms as

$$H \approx \frac{zN}{2}\left(\phi^2 + 3k\phi^4\right) - z\left(\phi + 2k\phi^3\right)\sum_{i=1}^{N} s_i. \tag{28}$$

As the Hamiltonian now only contains independent single-spin terms, the partition function becomes

$$Z = \sum_{\{s_i\}} \exp(-\beta H) = \exp\left[-\beta \frac{zN}{2}(\phi^2 + 3k\phi^4)\right] \prod_{i=1}^{N} \sum_{s_i = \pm 1} \exp\left[\beta z(\phi + 2k\phi^3) s_i\right]$$
$$= \exp\left[-\beta \frac{zN}{2}(\phi^2 + 3k\phi^4)\right] \left[2\cosh\left(\beta z(\phi + 2k\phi^3)\right)\right]^N \quad (29)$$

The mean-field free energy density is thus given by

$$f(\phi) = \frac{z}{2}\phi^2(1 + 3k\phi^2) - k_B T \ln\left[2\cosh\left(\beta z(\phi + 2k\phi^3)\right)\right]. \quad (30)$$

Expanding the free energy near $\phi = 0$ as a Taylor series yields

$$f(\phi) = f_0 + a(k,T)\phi^2 + b(k,T)\phi^4 + c(k,T)\phi^6 + O(\phi^7). \quad (31)$$

By analyzing the behavior of these coefficients, we obtain the phase diagram of the system. In particular, the sign change in the coefficient of the quartic term $b(k,T)$ indicates the presence of a tricritical point, where the nature of the phase transition changes from second-order to first-order as $k$ increases.

**Acknowledgments**

The computations of this work were conducted on the HPC cluster of ITP-CAS. Y.W. was partially supported by the National Natural Science Foundation of China (No.12047503) and Wenzhou Institute, University of Chinese Academy of Sciences (No. QD2023009).

**Figures and Tables**

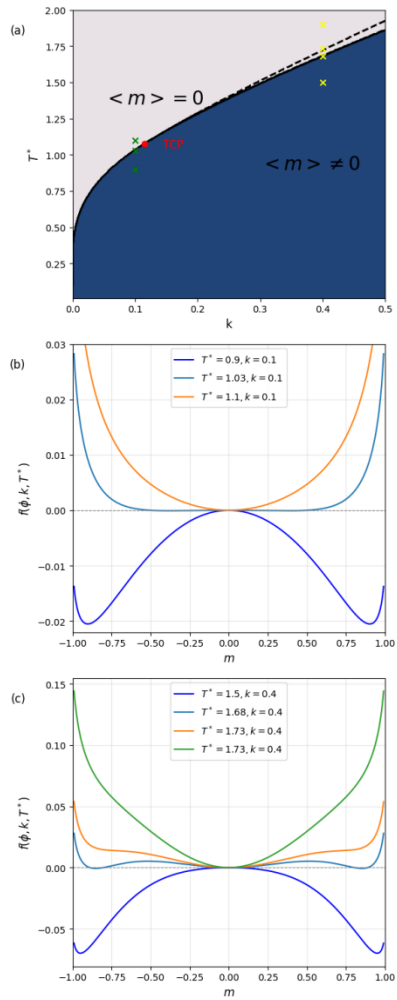

**Figure 1.** (a) Phase diagram derived from the exact one-dimensional solution with respect to the coupling parameter $k$ and the reduced temperature $T^*$. The white region indicates the paramagnetic phase $(\langle m \rangle = 0)$, the blue region represents the ferromagnetic phase $(\langle m \rangle \neq 0)$, and the dashed line shows the spinodal line. (b) Free energy landscape for $k = 0.1$, showing a continuous phase transition. (c) Free energy landscape for $k = 0.4$, showing a discontinuous phase transition.

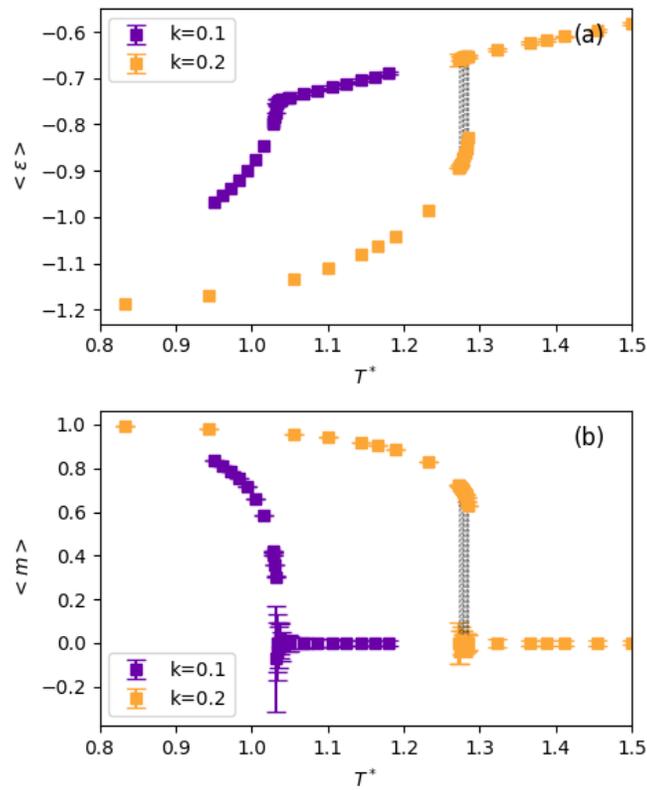

**Figure 2.** Temperature dependence of energy per spin (a) and magnetization per spin (b) for $k = 0.1$ and $k = 0.2$ from simulated data. For $k = 0.1$, both energy and magnetization change continuously with temperature, indicating a second-order phase transition. For $k = 0.2$, abrupt changes in energy and magnetization are observed near the transition temperature, suggesting a first-order phase transition.

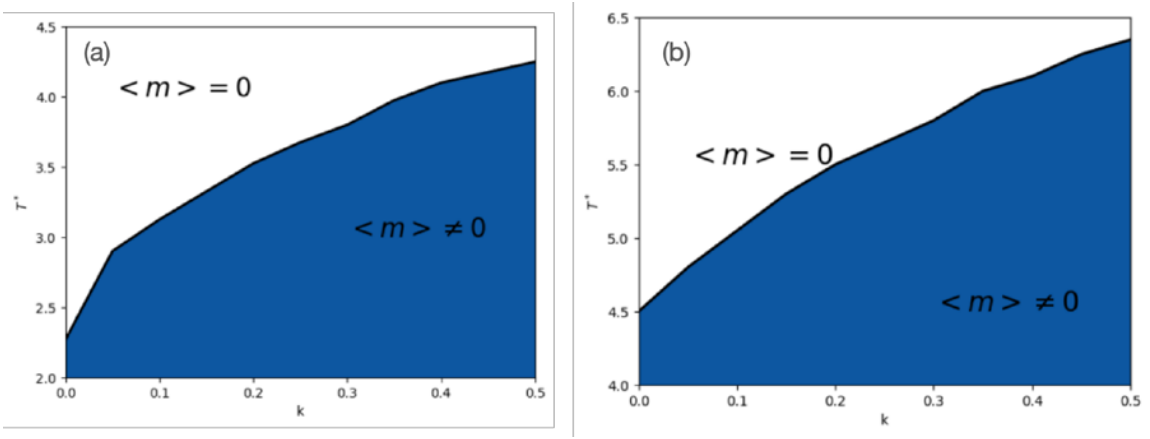

**Figure 3.** Phase diagrams of the 2D (a) and 3D (b) intelligent Ising models.

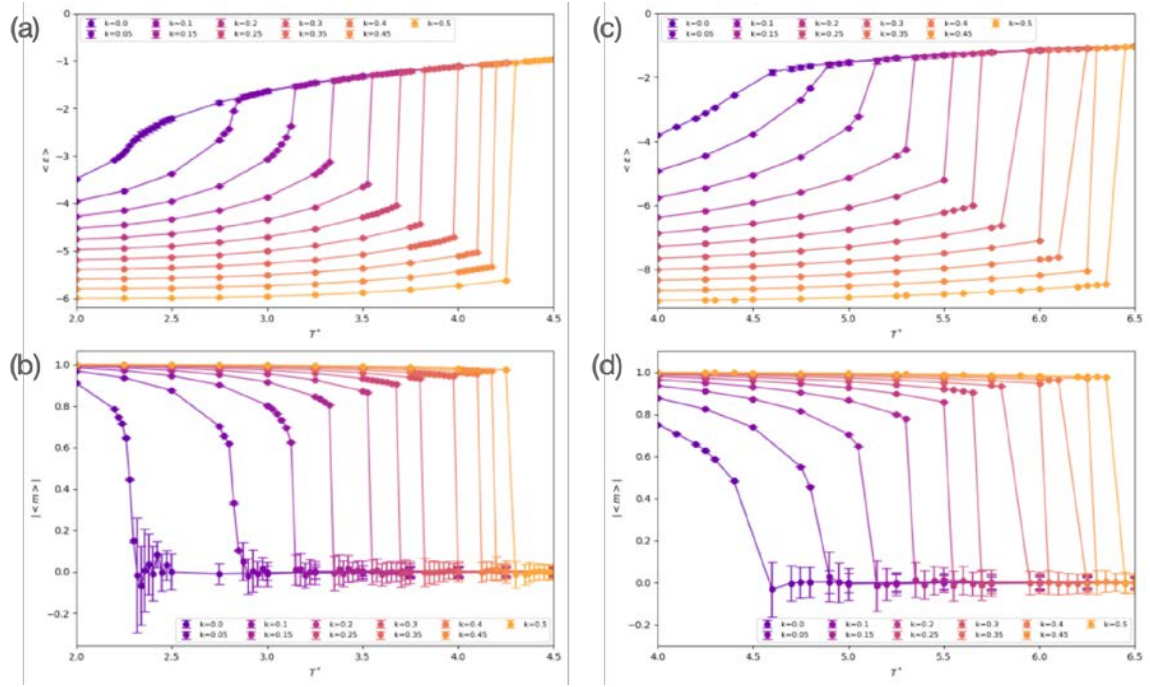

**Figure 4.** Temperature dependences of the energy per spin and the magnetization per spin for the 2D intelligent Ising model (a,b) and the 3D intelligent Ising model (c,d), with *k* varying from 0 to 0.5. The transitions become increasingly abrupt as *k* increases, and apparent discontinuities can be observed at larger *k* values.

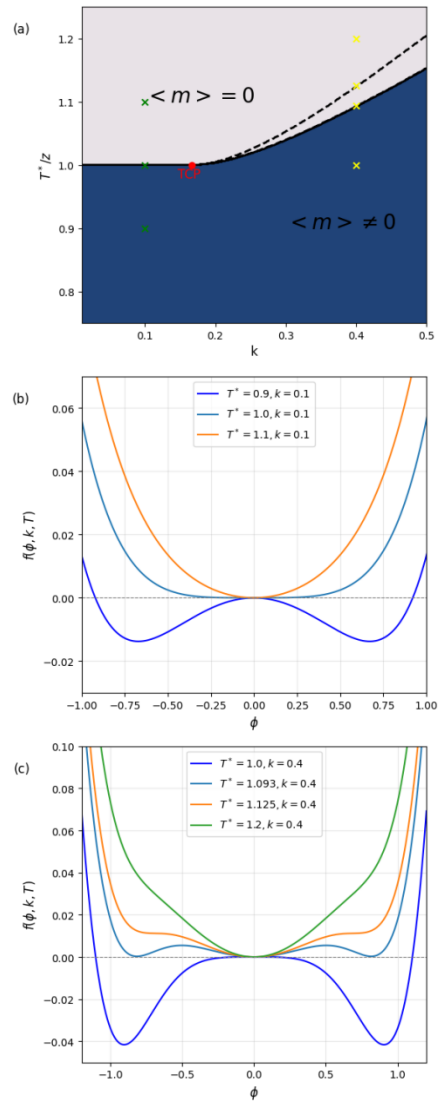

**Figure 5.** (a) Phase diagram derived from the mean-field theory with respect to the coupling parameter $k$ and the reduced temperature $T^*/z$. The white region indicates the paramagnetic phase $(\langle m \rangle = 0)$, the blue region represents the ferromagnetic phase $(\langle m \rangle \neq 0)$, and the dashed line shows the spinodal line. (b) Free energy landscape for $k = 0.1$, showing a continuous phase transition. (c) Free energy landscape for $k = 0.4$, showing a discontinuous phase transition.